# A Solution to Fastest Distributed Consensus Problem for Generic Star & K-cored Star Networks

Saber Jafarizadeh, *Student Member, IEEE,* and Abbas Jamalipour, *Fellow, IEEE*

*Abstract*—**Distributed average consensus is the main mechanism in algorithms for decentralized computation. In distributed average consensus algorithm each node has an initial state, and the goal is to compute the average of these initial states in every node. To accomplish this task, each node updates its state by a weighted average of its own and neighbors' states, by using local communication between neighboring nodes. In the networks with fixed topology, convergence rate of distributed average consensus algorithm depends on the choice of weights. This paper studies the weight optimization problem in distributed average consensus algorithm. The network topology considered here is a star network where the branches have different lengths. Closed-form formulas of optimal weights and convergence rate of algorithm are determined in terms of the network's topological parameters. Furthermore generic *K*-cored star topology has been introduced as an alternative to star topology. The introduced topology benefits from faster convergence rate compared to star topology. By simulation better performance of optimal weights compared to other common weighting methods has been proved.**

## I. INTRODUCTION

DISTRIBUTED consensus algorithm has received renewed interest due to its wide range of applications in on-line social networks, distributed databases and sensor networks.

A simple and common form of distributed consensus algorithm is distributed average consensus algorithm. In distributed average consensus algorithm, the nodes in the network have to compute the average of their initial states. Some of applications for distributed average consensus algorithm are gossip algorithms [1, 2], distributed estimation and detection [3, 4], sensor localization, [5], distributed data fusion in sensor networks, [6], multiagent distributed coordination and flocking [7, 8].

In distributed consensus averaging algorithm each node updates its state by a weighted average of its own and neighbors' states. Convergence rate of the algorithm depends on the choice of weights. A particular problem of interest is to find the optimal weights for distributed average consensus algorithm. This problem is known as the fastest distributed average consensus problem [9, 10].

S. Jafarizadeh and A. Jamalipour are with the School of Electrical and Information Engineering, University of Sydney, Sydney NSW 2006, Australia (e-mail: saber.jafarizadeh@sydney.edu.au, abbas.jamalipour@sydney.edu.au).

In this paper we have solved this problem for generic star network. By generic star network we refer to a simple star network where the branches have different lengths. We have provided closed formed formulas for the optimal weights and convergence rate of the algorithm. The main solution procedure comprises stratification and semidefinite programming. Stratification method reduces the number of variables in the semidefinite programming formulation. Meanwhile using the complementary slackness conditions we obtain the characteristic polynomials of weight matrix, which in turn results in optimal weights and convergence rate of algorithm. In addition generic *K*-cored star topology has been introduced as an alternative topology to generic star topology with more rapid convergence rate. This topology has been introduced as an option to overcome the slow convergence rate of generic star network caused by bottleneck effect of central node. Also by numerical simulations optimality of optimal weights has been confirmed.

The organization of the paper is as follows. Section II is a brief review of the literature in distributed average consensus algorithm. In Section III we describe distributed average consensus algorithm in detail. In Section IV we state our main results, including the evaluated optimal weights and convergence rate of algorithm. Section V is devoted to the proof of our main results. Section VI presents simulation results and we conclude the paper with discussion in Section VII.

## II. RELATED WORK

One of the pioneer works in distributed computation and consensus problem was [11] which had analyzed agreement algorithms in asynchronous and distributed environment.

Distributed consensus algorithm over sensor networks with time-varying topologies have been studied in [7, 12]. Authors in [7], have implemented a continuous time state update model for distributed consensus algorithm to deal with communication delay. Many works has studied distributed consensus in presence of noise [13, 14]. Kar and Moura [13] have studied average consensus algorithms in networks with random topologies and noisy communications. Their proposed algorithms reduce the mean and variance of error at the same time. Schizas et al [14] have considered distributed estimation scenario over ad hoc WSNs with quantization and noisy channels. Then they have modeled the resultant consensus problem as a multiple constrained convex optimization problems.

Several authors have studied distributed consensus algorithm with quantization constraints [15, 16]. Authors in [15] have modeled the quantization error as additive noise and they have reduced the variance of the quantization error by estimating the noise. In [16] Aysal et al. have used probabilistic quantization method for quantizing. They have shown that by using this method the convergence of nodes in assured but the average is not preserved.

References [17, 9, 10, 18, 19, 21] have addressed weight optimization problem in distributed consensus algorithm in networks with fixed topology. However in [19, 21] authors have solved this problem just for star networks with rich symmetric topological properties. In [9] Boyd et al determine the conditions over weight matrix for sure convergence of distributed average consensus algorithm. In addition they have formulated weight optimization problem in semidefinite programming formulation. In [10] the authors have used the symmetric properties of network to reduce the complexity of calculations. Here in this work we have solved the weight optimization problem for star network in its general form without any assumptions on network's topology.

## III. DISTRIBUTED AVERAGE CONSENSUS ALGORITHM

In this section we provide a brief review of distributed average consensus algorithm.

Distributed average consensus algorithm computes the average of initial states of nodes in a network. The term distributed initiates that the computation must be done by local communication between neighboring nodes. Distributed average consensus algorithm calculates the average ($\bar{x} = \mathbf{11}^T N x_0$) by updating the state of nodes according to the following iterative model.

$$x(t+1) = W \cdot x(t) \quad (1)$$

$x(0)$ is the column vector containing the initial states of nodes. $\mathbf{1}$ denotes the column vector with all elements one. $N$ is the number of nodes in the network. $x(t)$ is the column vector containing the states of nodes at time index $t = 1, 2, \ldots$. $W$ is the weight matrix with the same sparsity pattern as the adjacency matrix of network's associated connectivity graph. The necessary and sufficient conditions for the convergence of (1) are given in [9] as: (i) One is the single eigenvalue of $W$ associated with the eigenvector $\mathbf{1}$, and (ii) all other eigenvalues are strictly less than one in magnitude.

Finding the optimal weights ($W$) such that the iterative algorithm (1) converges with the fastest possible rate is known as fastest distributed average consensus problem. This problem can be formulated as the following minimization problem

$$\min_W \max(\lambda_2, -\lambda_N),$$
$$s.t. \quad W = W^T, W \cdot \mathbf{1} = \mathbf{1}, \forall \{i, j\} \notin \mathcal{E}: W_{ij} = 0,$$

where $\lambda_i$ for $i = 1, \ldots, N$ are the eigenvalues of $W$ arranged in decreasing order ($\lambda_1 = 1$). The term $\max(\lambda_2, -\lambda_N)$ is the *Second Largest Eigenvalue Modulus* (*SLEM*) of $W$. FDC averaging problem can be formulated in the semidefinite programming form as [9]:

$$\min_W \quad s$$
$$s.t. \quad -sI \preceq W - \mathbf{11}^T/N \preceq sI, W = W^T \quad (2)$$
$$W \cdot \mathbf{1} = \mathbf{1}, \forall \{i, j\} \notin \mathcal{E}: W_{ij} = 0.$$

We refer to problem (2) as the fastest distributed average consensus problem.

## IV. MAIN RESULTS

In this section we state our main results concerning optimal weights and SLEM of distributed average consensus algorithm over generic star and generic $K$-cored star networks. Complete proof of our results is presented in section V.

### A. Generic Star Topology

A generic star network consists several path branches connected to one central node. Here we have considered a star network with $B$ different types of path branches (each with different length). $n_i$ ($i$-th element of vector $n$) is the number of branches with length $m_i$ ($i$-th element of vector $m$) where $i$ varies from 1 to $B$. By length of each branch we refer to the number of nodes in each branch. A generic star graph with $B = 3$, $n = [4 \quad 3 \quad 3]$, $m = [1 \quad 2 \quad 3]$ is depicted in Fig. 1.

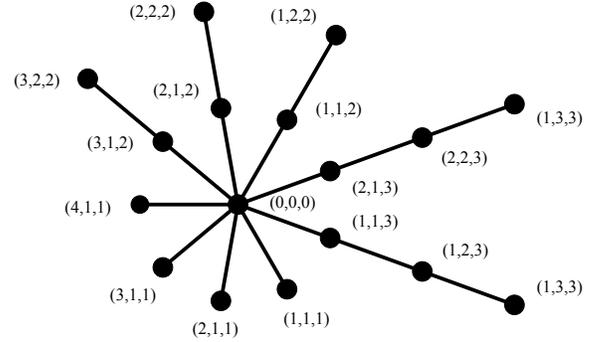

Fig. 1. A generic star network with $B = 3$, $n = [4 \quad 3 \quad 3]$, $m = [1 \quad 2 \quad 3]$.

The optimal weights for all of the edges on star network equal $1/2$ except those connecting path branches to the central node. The optimal weights for the edges connecting branches to the central node are as follows.

$$w_1^{(p)} = \frac{(1 - \cos(\theta)) \sin(m_p \theta)}{\sin(m_p \theta) - \sin((m_p - 1)\theta)}. \quad (3)$$

Equation (3) holds for $p = 1, \ldots, B$. $\theta$ is the smallest root of $|A| = 0$ in the interval $(0, \pi)$, where $|A|$ is the determinant of matrix $A$ defined as below

$$A_{i,j} = \begin{cases} \dfrac{2}{n_i} \cot(m_i \theta) \cot\left(\dfrac{\theta}{2}\right) - 1 & \text{for } i = j = 1, \ldots, B, \\ -\sqrt{\dfrac{n_j}{n_i}} & \text{for } i \neq j. \end{cases} \quad (4)$$

The Second Largest Eigenvalue Modulus (SLEM) of generic star network equals $\cos(\theta)$. The SLEM of generic star network with six branches of three different lengths ($B = 3$) are displayed in Table 1. These results are obtained for the choice of optimal weights (3).

Table 1. *SLEM* of generic star network for different combination of length ($m$) and number ($n$) of branches.

| $m \backslash n$ | $\begin{bmatrix}1\\2\\3\end{bmatrix}$ | $\begin{bmatrix}3\\2\\1\end{bmatrix}$ | $\begin{bmatrix}3\\1\\2\end{bmatrix}$ | $\begin{bmatrix}1\\3\\2\end{bmatrix}$ | $\begin{bmatrix}2\\3\\1\end{bmatrix}$ | $\begin{bmatrix}2\\1\\3\end{bmatrix}$ |
|---|---|---|---|---|---|---|
| [3 2 1] | 0.8990 | 0.9223 | 0.9200 | 0.9025 | 0.9157 | 0.9102 |
| [4 2 1] | 0.9352 | 0.9483 | 0.9477 | 0.9358 | 0.9434 | 0.9421 |
| [4 3 1] | 0.9378 | 0.9505 | 0.9488 | 0.9401 | 0.9472 | 0.9433 |
| [4 3 2] | 0.9402 | 0.9512 | 0.9501 | 0.9418 | 0.9479 | 0.9454 |
| [5 3 1] | 0.9569 | 0.9645 | 0.9639 | 0.9574 | 0.9617 | 0.9606 |
| [5 3 2] | 0.9575 | 0.9647 | 0.9644 | 0.9578 | 0.9620 | 0.9613 |
| [5 4 1] | 0.9582 | 0.9657 | 0.9645 | 0.9596 | 0.9638 | 0.9612 |
| [5 4 2] | 0.9589 | 0.9659 | 0.9650 | 0.9602 | 0.9641 | 0.9619 |
| [5 4 3] | 0.9602 | 0.9663 | 0.9657 | 0.9611 | 0.9644 | 0.9630 |

From Table 1 it is obvious that increasing either the length or the number of branches in generic star topology slows down the convergence rate. Also SLEM of network is more affected by increasing length or number of longer branches.

**Remark 1**: Two special cases of a star network with $B = 1$ and $B = 2$ has been studied in references [19] and [18], respectively. For $B = 1$ the star network reduces to a symmetric star network and the equation $|A| = 0$ results in

$$(n-2)\cos\left(\left(m-\frac{1}{2}\right)\theta\right) = (n+2)\cos\left(\left(m+\frac{1}{2}\right)\theta\right),$$

which is the same result obtained in [19]. In the case of $B = 2$ the star network reduces to a two fused star network and the equation $|A| = 0$ reduces to

$$\left(\frac{2}{n_1}\cot(m_1\theta)\cot\left(\frac{\theta}{2}\right) - 1\right) \times \left(\frac{2}{n_2}\cot(m_2\theta)\cot\left(\frac{\theta}{2}\right) - 1\right) = 1$$

which is the same exact result obtained in [18].

**Remark 2**: For the choice of optimal weights the following eigenvalues have the same absolute value.
- Largest eigenvalue of matrices obtained from stratification (5-a),
- Second largest eigenvalue of $W_0$ (given in (5-b)),
- Smallest eigenvalues of $W_0$.

It should be mentioned that one is the largest eigenvalue of $W_0$. Another important issue is that the optimal weights for the edges connecting shorter branches to central node are smaller than the ones connecting longer branches. Because it takes more number of iterations to mix the information in longer branches compared to shorter ones. But this is not true for the weights obtained from other weighting methods, namely, maximum degree [9], Metropolis-Hasting [12] and best constant [20] (Appendix C) methods (these methods are commonly used in the literature). This is due to the fact that using these weighting methods; same weight is assigned to all of the edges connected to central node. This conclusion holds true for all of the nodes which are acting as a bottleneck in other topologies.

In the generic star topology, central node acts as the bottleneck of network and slows down the convergence rate of average consensus algorithm. A simple way to reduce the bottleneck effect and enhance the convergence rate is to add new nodes parallel with the central one. The resultant topology in its general form is generic $K$-cored star topology (see Fig. 2). In generic $K$-cored star topology for adding each parallel central node, $\sum_{i=1}^{B} n_i$ new connections are required.

### B. Generic K-Cored Star Topology

A generic $K$-cored star network is a generic star network with $K$ parallel nodes at center. We define generic $K$-cored star topology by four parameters: scalars $B$ and $K$ and vectors $n$ and $m$. In generic $K$-cored star topology, $B$ different types of branches (each with different length) are connected to $K$ parallel central nodes and $n_i$ determines the number of branches with length $m_i$, ($i$ varies from 1 to $B$). A generic $K$-cored star graph with $B = 3$, $K = 2$, $n = [3\ 2\ 2]$, $m = [2\ 3\ 4]$ is depicted in Fig. 2.

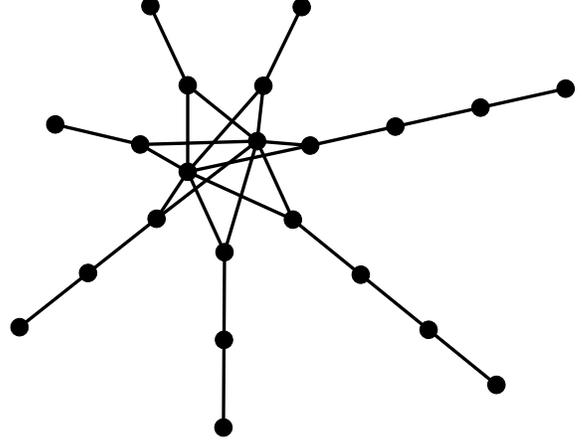

Fig. 2. A generic $K$-cored star network with $K = 2$, $B = 3$, $n = [3\ 2\ 2]$, $m = [2\ 3\ 4]$.

The optimal weights for all of the edges on generic $K$-cored star network equal $1/2$ except those connecting branches to the parallel central nodes. The optimal weights for the edges connecting branches to the central nodes are the same as (3) normalized by $K$. Second Largest Eigenvalue Modulus (SLEM) of generic $K$-cored star network equals $\cos(\theta)$. $\theta$ is the smallest root of $|A| = 0$ in the interval $(0, \pi)$, where $A$ is the same matrix defined at (4) but with $(2K/n_i)\cot(m_i\theta)\cot(\theta/2) - 1$ as the diagonal entries.

**Remark 3.** The optimal weights for generic $K$-cored star network are obtained by ignoring the single eigenvalue $1 - \sum_{p=1}^{B} n_p w_1^p$. Therefore these results are true for the values of $K$ where $1 - \sum_{p=1}^{B} n_p w_1^p$ is smaller than SLEM. The values of $K$ satisfying this constraint are between one and $K_{max}$, where $K_{max}$ for a few combination of branches is presented in Table 2. The results in Table 2 are obtained numerically.

Table 2. $K_{max}$ in terms of length ($m$) and number ($n$) of branches

| $m \backslash n$ | $\begin{bmatrix}1\\2\\3\end{bmatrix}$ | $\begin{bmatrix}3\\2\\1\end{bmatrix}$ | $\begin{bmatrix}3\\1\\2\end{bmatrix}$ | $\begin{bmatrix}1\\3\\2\end{bmatrix}$ | $\begin{bmatrix}2\\3\\1\end{bmatrix}$ | $\begin{bmatrix}2\\1\\3\end{bmatrix}$ |
|---|---|---|---|---|---|---|
| [3 2 1] | 15 | 27 | 25 | 16 | 22 | 19 |
| [4 2 1] | 20 | 43 | 42 | 21 | 33 | 30 |
| [4 3 1] | 24 | 47 | 44 | 27 | 39 | 32 |
| [4 3 2] | 28 | 49 | 47 | 30 | 40 | 36 |
| [5 3 1] | 30 | 68 | 65 | 33 | 52 | 46 |
| [5 3 2] | 33 | 69 | 68 | 35 | 53 | 50 |
| [5 4 1] | 36 | 74 | 68 | 42 | 61 | 49 |
| [5 4 2] | 39 | 75 | 71 | 44 | 62 | 53 |
| [5 4 3] | 44 | 77 | 74 | 47 | 64 | 57 |

**Remark 4.** SLEM of generic $K$-cored star network increases as $K$ gets larger than $K_{max}$. In Fig. 3 SLEM of the network illustrated in Fig. 2 is depicted in terms of $K$ (number of parallel central nodes). The optimal values of weights are calculated numerically. As it is obvious from Fig. 3 minimum value of SLEM is obtained for $K$ equal to $K_{max}$.

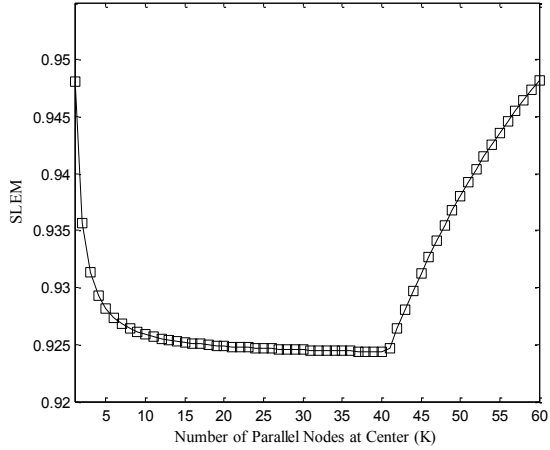

Fig. 3. SLEM of generic $K$-cored star topology with $B = 3$, $\boldsymbol{n} = [3 \ 2 \ 2]$, $\boldsymbol{m} = [2 \ 3 \ 4]$ in terms of number of parallel nodes at center ($K$).

### C. Generic Star & Generic K-Cored Star, Networks with Branches other than Path

In topologies introduced in this section, the type of branches is not limited to path graph. In [21] four branches other than path branch are introduced with their corresponding optimal weights. These branches are Lollipop, Semi-Complete, Ladder and Palm branches. It has been proved that the optimal weights presented in [21] are independent of the rest of network. These branches can be used in both configurations described previously in this section, while their optimal weights would remain the same. In the following, we have provided an example of such topologies.

*Example.* In Fig. 4 a generic $K$-cores star topology with two parallel central nodes and a combination of different branches is depicted. The optimal weights are as illustrated in Fig. 4. SLEM of this topology is 0.96551.

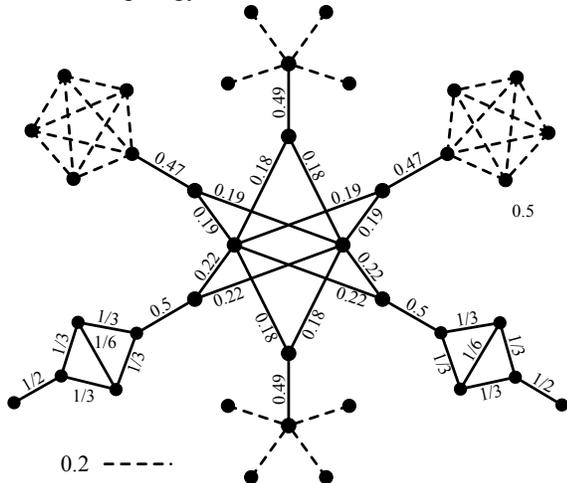

Fig. 4. a generic $K$-cores star topology with two parallel central nodes, two lollipop branches of order (5,1), two palm branches of order (1,4) and two semi-complete branch of order (4,1,1).

## V. PROOF OF MAIN RESULTS

In this section we provide the solution of fastest distributed average consensus algorithm for a network with generic star topology.

*Generic Star Topology*

Here we have considered a generic star network as described in section IV-A. We model the topology of generic star network (established communication channels between nodes) by undirected graph $\mathcal{G} = (\mathcal{V}, \mathcal{E})$. Graph $\mathcal{G}$ consists $|\mathcal{V}| = 1 + \sum_{i=1}^{B} m_i n_i$ nodes and $|\mathcal{E}| = \sum_{i=1}^{B} m_i n_i$ edges (see Fig. 1 for $B = 3$, $\boldsymbol{n} = [4 \ 3 \ 3]$, $\boldsymbol{m} = [1 \ 2 \ 3]$). The set of nodes is $\mathcal{V} = \{(i,j,p): p = 1, ..., B, \ i = 1, ..., n_p, \ j = 1, ..., m_p\} \cup \{(0,0,0)\}$, where (0,0,0) denotes the central node.

The automorphism group of generic star graph $Aut(\mathcal{G})$ is isomorphic to permutation of branches with the same length $(S_{n_1} \otimes S_{n_2} \otimes ... \otimes S_{n_B})$. Thus we can state that generic star graph has $1 + \sum_{i=1}^{B} m_i$ orbits acting on vertices and $\sum_{i=1}^{B} m_i$ edge orbits. The orbits of $Aut(\mathcal{G})$ acting on the vertices are

$$\{(0,0,0)\} \cup \left\{(1,j,p), (2,j,p), ..., (n_p, j, p) \ \middle| \ \begin{matrix} j = 1, ..., m_p, \\ p = 1, ..., B \end{matrix}\right\}.$$

The edge orbits of $Aut(G)$ are

$$\left\{((0,0,0), (i,1,p)) \ \middle| \ \begin{matrix} i = 1, ..., n_p \\ p = 1, ..., B \end{matrix}\right\},$$

and

$$\{((i,j-1,p), (i,j,p)) | i = 1, ..., n_p\} \text{ for } \begin{matrix} j = 2, ..., m_p, \\ p = 1, ..., B. \end{matrix}$$

*Stratification of Star Topology*

For a review of stratification method and derivation of semidefinite programming used in this section we refer to [18].

The optimal weights on the edges within an edge orbit (stratum) are the same [10]. Thus it suffices to consider $\sum_{i=1}^{B} m_i$ weights, namely, $w_1^{(p)}, w_2^{(p)}, ..., w_{m_p}^{(p)}$ where $p$ vary from 1 to $B$. We associate with each node $(i,j,p)$, the $|\mathcal{V}| \times 1$ column vector

$$\boldsymbol{e}_{i,j}^{(p)} = \boldsymbol{d}_i \otimes \boldsymbol{f}_j \otimes \boldsymbol{h}_p \text{ for } \begin{matrix} i = 1, ..., n_p \\ j = 1, ..., m_p \\ p = 1, ..., B \end{matrix}$$

$\boldsymbol{d}_i, \boldsymbol{f}_j$ and $\boldsymbol{h}_p$ are $B \times 1$ and $n_p \times 1$ and $m_p \times 1$ column vectors with one in the $i$-th, $j$-th and $p$-th position respectively and zero elsewhere. We denote the central node (0,0,0) by $\boldsymbol{e}_{0,0}^{(0)}$ with one in the last position and zeros elsewhere. Based on the above vector space, we define the weight matrix as following.

$$\boldsymbol{W}_{(i,j,p),(\mu,\eta,q)} = \begin{cases} w_1^q & \text{if } i = j = p = 0, \ \mu \in [1, n_q], \eta = 1, q \in [1, B] \\ w_\eta^p & \text{if } i = \mu \in [1, n_p], p = q \in [1, B], j+1 = \eta \in [2, m_p] \\ 1 - \sum_{p=1}^{B} n_p w_1^{(p)} & \text{if } i = j = p = \mu = \eta = q = 0 \\ 1 - w_j^{(p)} - w_{j+1}^{(p)} & \text{if } i = \mu \in [1, n_p], p = q \in [1, B], j = \eta \in [1, n_p] \end{cases}$$

The definition above ensures that the weight matrix has an eigenvalue equal to one and the corresponding eigenvector is a vector with all elements equal to one.

Using the unitary DFT matrix of size $n_p \times n_p$ for $p = 1, \ldots, B$ in each edge stratum, a new set of orthonormal vectors are defined as follows.

$$\varphi_{k,\mu}^p = \begin{cases} \frac{1}{\sqrt{n_p}} \sum_{i=1}^{n_p} \omega_p^{\mu(i-1)} e_{i,k}^{(p)} & \text{for } \begin{array}{l} p = 1, \ldots, B, \\ k = 1, \ldots, m_p, \\ \mu = 0, \ldots, n_p - 1, \end{array} \\ e_{0,0}^0 & \text{for } i = \mu = p = 0. \end{cases}$$

$\omega_p$ equals $exp(j2\pi/n_p)$ for $p = 1, \ldots, B$ ($j$ is the imaginary unit). The weight matrix in the new basis takes the following block diagonal form

$$W = diag(W_0, W_1, \ldots, W_B).$$

The Matrices $W_p$ for $p = 1, \ldots, B$ are as follows.

$$W_p = \begin{bmatrix} 1 - w_1^{(p)} - w_2^{(p)} & w_2^{(p)} & 0 & \cdots \\ w_2^{(p)} & 1 - w_2^{(p)} - w_3^{(p)} & w_3^{(p)} & \ddots \\ 0 & w_3^{(p)} & \ddots & w_{m_p}^{(p)} \\ \vdots & \ddots & w_{m_p}^{(p)} & 1 - w_{m_p}^{(p)} \end{bmatrix}.$$

(5-a)

For the matrix $W_0$ we have

$$W_0 = \begin{bmatrix} 1 - \sum_{i=1}^{B} n_i w_1^{(i)} & \sqrt{n_1} w_1^{(1)} h_1^T & \sqrt{n_2} w_1^{(2)} h_2^T & \cdots & \sqrt{n_B} w_1^{(B)} h_B^T \\ \sqrt{n_1} w_1^{(1)} h_1 & W_1 & 0 & \ddots & 0 \\ \sqrt{n_2} w_1^{(2)} h_2 & 0 & W_2 & \ddots & \vdots \\ \vdots & \ddots & \ddots & \ddots & 0 \\ \sqrt{n_B} w_1^{(B)} h_B & 0 & \cdots & 0 & W_B \end{bmatrix},$$

(5-b)

where $h_p$ for $p = 1, \ldots, B$ is a $m_p \times 1$ column vector with one in first position and zero elsewhere. Introducing $W_0'$ as

$$W_0' = diag(W_1, W_2 \ldots, W_B), \quad (6)$$

and considering the fact that $W_0'$ is a submatrix of $W_0$, by using *Cauchy Interlacing Theorem* (provided in Appendix A), we can state the following corollary for the eigenvalues of $W_0$ and $W_0'$.

*Corollary 1,*
If we consider $W_0$ and $W_0'$ as in (5-b) and (6), respectively, then theorem 1 implies the following relations between the eigenvalues of $W_0$ and $W_0'$,

$$\lambda_{1+\sum_{p=1}^{B} m_p}(W_0) \leq \lambda_{\sum_{p=1}^{B} m_p}(W_0') \leq \cdots \leq \lambda_2(W_0') \leq \lambda_2(W_0)$$
$$\leq \lambda_1(W_0') \leq \lambda_1(W_0) = 1$$

It is obvious from above relations that second largest eigenvalue $(\lambda_2(W))$ of weight matrix is the largest eigenvalue of $W_0'$, while smallest eigenvalue $\left(\lambda_{1+\sum_{i=1}^{B} m_i n_i}(W)\right)$ of weight matrix is the smallest eigenvalue of $W_0$.

In the case of $n_1 = 1$, the weight matrix $W$ does not include $W_1$ and consequently difference between dimensions of $W_0$ and $W_0'$ will be more than one and corollary 1 will not stand true. It is clear that the same is true for $n_p = 1$ and $W_p$ for $p = 2, \ldots, B$, thus corollary 1 is true for $n_1, n_2, \ldots, n_B \geq 2$.

*Determination of Optimal Weights via Semidefinite Programming*

Based on corollary 1, and section III, fastest distributed average consensus problem for a network with generic star topology can be formulated as the following semidefinite programming problem.

$$\begin{aligned} \min \quad & s \\ s.t. \quad & W_0' \leq sI_{\sum_{p=1}^{B} m_p}, \\ & -sI_{1+\sum_{p=1}^{B} m_p} \leq W_0 - vv^T. \end{aligned} \quad (7)$$

$v$ is the eigenvector of $W_0$ corresponding to the eigenvalue one. It is a $(M_B + 1) \times 1$ column vector defined as:

$$v(i) = \frac{1}{\sqrt{1 + \sum_{p=1}^{B} m_p n_p}} \times \begin{cases} 1 & \text{for } i = 1, \\ \sqrt{n_p} & \text{for } \begin{array}{l} i = 2 + M_{p-1}, \ldots, M_p + 1, \\ p = 1, \ldots, B, \end{array} \end{cases}$$

where $M_p$ is defined as $\sum_{j=1}^{p} m_j$ for $p = 1, \ldots, B$.

$W_0$ and $W_0'$ can be written as a linear combination of rank one matrices as following,

$$W_0' = I - \sum_{p=1}^{B} \sum_{i=1}^{m_p} w_i^{(p)} \alpha_i^{(p)} \alpha_i^{(p)T}, \quad (8\text{-a})$$

$$W_0 = I - \sum_{p=1}^{B} \sum_{i=1}^{m_p} w_i^{(p)} \beta_i^{(p)} \beta_i^{(p)T}, \quad (8\text{-b})$$

The vectors $\alpha_i^{(p)}$ and $\beta_i^{(p)}$ are provided in appendix B.
Using decompositions (8), the constraints in (7) can be written as

$$sI - I + \sum_{p=1}^{B} \sum_{i=1}^{m_p} w_i^{(p)} \alpha_i^{(p)} \alpha_i^{(p)T} \geq 0 \quad (9\text{-a})$$

$$sI + I - vv^T - \sum_{p=1}^{B} \sum_{i=1}^{m_p} w_i^{(p)} \beta_i^{(p)} \beta_i^{(p)T} \geq 0 \quad (9\text{-b})$$

In the following we formulate problem (7) in the form of standard semidefinite programming (described in [9]). Problem parameters ($F_i$, $c$) are defined as

$$F_0 = \begin{bmatrix} -I_{M_B \times M_B} & 0 \\ 0 & I_{(1+M_B)\times(1+M_B)} - vv^T \end{bmatrix}$$

$$F_{i,p} = \begin{bmatrix} \alpha_i^{(p)} \alpha_i^{(p)T} & 0 \\ 0 & -\beta_i^{(p)} \beta_i^{(p)T} \end{bmatrix} \text{ for } \begin{array}{l} p = 1, \ldots, B \\ i = 1, \ldots, m_p \end{array}$$

$$F_{1+M_B} = I_{1+2M_B},$$

$$c_i = 0, \quad i = 1, \ldots, M_B, \quad c_{1+M_B} = 1,$$

Minimization variable ($x$) is defined as

$$x^T = \left[ w_1^{(1)}, w_2^{(1)}, \ldots, w_{m_1}^{(1)}, w_1^{(2)}, w_2^{(2)}, \ldots, w_{m_B}^{(B)}, s \right]$$

In the case of dual problem we choose the dual variable $Z$ as $Z = \begin{bmatrix} z_1 \\ z_2 \end{bmatrix} \begin{bmatrix} z_1^T & z_2^T \end{bmatrix}$ to ensure that $Z$ is positive definite. $z_1$ and $z_2$ are column vectors, with $M_B$ and $M_B + 1$ elements, respectively. From the constraints of dual problem we obtain:

$$\left( \alpha_i^{(p)T} z_1 \right)^2 = \left( \beta_i^{(p)T} z_2 \right)^2, \text{ for } \begin{array}{l} p = 1, \ldots, B \\ i = 1, \ldots, m_p \end{array}, \quad (10)$$

Complementary slackness condition ($F(x)Z = ZF(x) = 0$) [18] dictates the following relation for optimal values of primal feasible point ($x$) and dual feasible point ($Z$),

$$(sI - W'_0)z_1 = 0, \quad (sI + W_0 - vv^T)z_2 = 0. \quad (11)$$

Multiplying both sides of (11) by $vv^T$ we have $svv^T z_2 = 0$ which implies that $v^T z_2 = 0$. Consequently (11) reduces to

$$(sI - W'_0)z_1 = 0, \quad (sI + W_0)z_2 = 0. \quad (12)$$

Since the vectors $\alpha_i^{(p)}$ and $\beta_i^{(p)}$ form a basis for their corresponding vector spaces, we can expand $z_1$ and $z_2$ in terms of $\alpha_i^{(p)}$ and $\beta_i^{(p)}$ as following

$$z_1 = \sum_{p=1}^{B} \sum_{i=1}^{m_p} a_i^{(p)} \alpha_i^{(p)}, \quad z_2 = \sum_{p=1}^{B} \sum_{i=1}^{m_p} b_i^{(p)} \beta_i^{(p)}, \quad (13)$$

with the coordinates $a_i^{(p)}$ and $b_i^{(p)}$ for $i = 1, \ldots, m_p$, $p = 1, \ldots, B$ to be determined.

Using the expansions (8) and (13), by comparing the coefficients of $\alpha_i^{(p)}$ and $\beta_i^{(p)}$ in slackness conditions (12), we obtain

$$(-s + 1)a_i^{(p)} = w_i^{(p)} \left( \alpha_i^{(p)T} z_1 \right), \quad (14\text{-a})$$

$$(s + 1)b_i^{(p)} = w_i^{(p)} \left( \beta_i^{(p)T} z_2 \right), \quad (14\text{-b})$$

(14) hold for $p = 1, \ldots, B$ and $i = 1, \ldots, m_p$.

From (14) and dual constraints (10), we can deduce that $(s + 1)^2 \left( a_i^{(p)} \right)^2 = (-s + 1)^2 \left( b_i^{(p)} \right)^2$ for $p = 1, \ldots, B$ and $i = 1, \ldots, m_p$, which is equivalent to

$$\left( a_i^{(p)} / a_j^{(q)} \right)^2 = \left( b_i^{(p)} / b_j^{(q)} \right)^2 : \begin{array}{l} \forall i, j \in [1, m_p], \\ \forall p, q = [1, B], \end{array} \quad (15)$$

By expressing $\left( \alpha_i^{(p)T} z_1 \right)$ and $\left( \beta_i^{(p)T} z_2 \right)$ in terms of the coefficients $a_i^{(p)}$ and $b_i^{(p)}$, the equations (14) can be reduced to the following recursive equations,

$$\left( -s + 1 - w_1^{(p)} \right) a_1^{(p)} = -w_1^{(p)} a_2^{(p)} \quad (16\text{-a})$$

$$\left( -s + 1 - 2w_i^{(p)} \right) a_i^{(p)} = -w_i^{(p)} \left( a_{i-1}^{(p)} + a_{i+1}^{(p)} \right), \quad (16\text{-b})$$

$$\left( -s + 1 - 2w_{m_p}^{(p)} \right) a_{m_p}^{(p)} = -w_{m_p}^{(p)} a_{m_p-1}^{(p)} \quad (16\text{-c})$$

and

$$\left( s + 1 - (1 + n_p) w_1^{(p)} \right) b_1^{(p)} - w_1^{(p)} \sum_{\substack{j=1 \\ j \neq p}}^{B} \sqrt{n_p n_j} b_1^{(j)} = -w_1^{(p)} b_2^{(p)} \quad (17\text{-a})$$

$$\left( s + 1 - 2w_i^{(p)} \right) b_i^{(p)} = -w_i^{(p)} \left( b_{i-1}^{(p)} + b_{i+1}^{(p)} \right) \quad (17\text{-b})$$

$$\left( s + 1 - 2w_{m_p}^{(p)} \right) b_{m_p}^{(p)} = -w_{m_p}^{(p)} b_{m_p-1}^{(p)} \quad (17\text{-c})$$

where $p$ varies from 1 to $B$. (16-b) and (17-b) hold for $i = 1, \ldots, m_p$.

Now we can determine $s$ (*SLEM*), the optimal weights and the coordinates $a_i^{(p)}$ and $b_i^{(p)}$, in an inductive manner as follows.

In the first stage, from comparing equations (16-c) and (17-c) and considering the relation (15), we achieve

$$\left( -s + 1 - 2w_{m_p}^{(p)} \right)^2 = \left( s + 1 - 2w_{m_p}^{(p)} \right)^2,$$

which results in $w_{m_p}^{(p)} = 1/2$. Assuming $s = \cos(\theta)$ and substituting $w_{m_p}^{(p)} = 1/2$ in (16-c) and (17-c), we have

$$a_{m_p-1}^{(p)} = \frac{\sin(2\theta)}{\sin(\theta)} a_{m_p}^{(p)}, \quad b_{m_p-1}^{(p)} = \frac{\sin(2(\pi - \theta))}{\sin(\pi - \theta)} b_{m_p}^{(p)}$$

Continuing the above procedure inductively, up to $i - 1$ stages, we achieve

$$w_i = 1/2, \qquad (18)$$

and

$$a_{i-1}^{(p)} = \left(\frac{\sin\left((\boldsymbol{m}_p - i + 2)\theta\right)}{\sin(\theta)}\right) a_{\boldsymbol{m}_p}^{(p)}, \qquad (19\text{-a})$$

$$b_{i-1}^{(p)} = \left(\frac{\sin\left((\boldsymbol{m}_p - i + 2)(\pi - \theta)\right)}{\sin(\pi - \theta)}\right) b_{\boldsymbol{m}_p}^{(p)}. \qquad (19\text{-b})$$

The results in (18) and (19) are true for $i = 2, \dots, \boldsymbol{m}_p$ and $p = 1, \dots, B$.

Using equations (19), $a_1^{(p)}, a_2^{(p)}$ and $b_1^{(p)}, b_2^{(p)}$ can be expressed in terms of $a_{\boldsymbol{m}_p}^{(p)}$ and $b_{\boldsymbol{m}_p}^{(p)}$ for $p = 1, \dots, B$. Then substituting the results in equations (16-a) we can find the optimal weights $w_1^{(p)}$ in terms of $\theta$ as given in (3). By substituting (3) in (17-a) and solving the equations in terms of $\theta$, one can find out $\theta$ as the smallest root of $|\boldsymbol{A}| = 0$ in the interval $(0, \pi)$ where matrix $\boldsymbol{A}$ is defined as (4).

## VI. SIMULATION RESULTS

This section includes numerical simulations comparing the performance of optimal weights with other weighting methods which are commonly used in the literature. The weighting methods we have considered are maximum degree [9], Metropolis-Hasting [12] and best constant [20] weighting methods (Appendix C). The topology we have considered for our simulation is the star network depicted in Fig. 1.

The asymptotic convergence rate of distributed average consensus is inversely proportional to the SLEM of network. To have a comparison between weighting methods based on asymptotic convergence rate, In Table 3 we have provided the SLEM of weight matrix with different weights. Furthermore in Fig. 5 we have investigated per step convergence rate of average consensus algorithm for different weighting methods. In Fig. 5 Euclidean distance of vector of node values $\boldsymbol{x}(t)$ from the mean value, in terms of number of iterations is presented. It should be mentioned that the results depicted in Fig. 5 are in logarithmic scale and generated based on 10000 trials (a different random initial node values is generated for each trial).

From the results presented in Table 3 and Fig. 5 it is obvious that the optimal weights have better performance compared to other three weights.

Table 3. SLEM of star network with $B = 3$, $\boldsymbol{n} = [4\ \ 3\ \ 2]$, $\boldsymbol{m} = [1\ \ 2\ \ 3]$, for different weighting methods.

| Weighting Method | Metropolis | Maximum Degree | Best Constant | Optimal |
|---|---|---|---|---|
| SLEM | 0.9718 | 0.9780 | 0.9614 | 0.9213 |

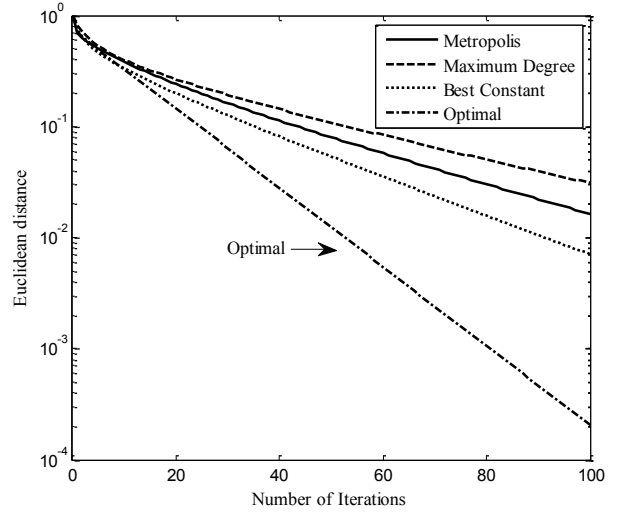

Fig. 5. Normalized Euclidean Distance of vector of node values $\boldsymbol{x}(t)$ from the mean value in terms of number of iterations for a star network with $B = 3$, $\boldsymbol{n} = [4\ \ 3\ \ 2]$, $\boldsymbol{m} = [1\ \ 2\ \ 3]$ and different weighting methods.

## VII. CONCLUSION

In this paper we have considered weight optimization problem in distributed average consensus algorithm. We have solved this problem for a network with a star topology where branches have different lengths. Also we have introduced generic $K$-cored star topology which has more rapid convergence rate compared to star network. Using stratification method we convert the weight matrix into block diagonal form. Then by enforcing the complementary slackness conditions we reach the characteristic polynomials of weight matrix. Solving these polynomials in an inductive manner leads to the optimal weights and SLEM of network. Examples of star network with branches other than path branches are provided. By Simulation we verify the better performance of optimal weights compared to other common weighting methods. Our future directions include the addition of noise and communication delay in asynchronous mode.

## APPENDIX A

### THEOREM 1 (CAUCHY INTERLACING THEOREM) [22]

Let $\boldsymbol{A}$ and $\boldsymbol{B}$ be $n \times n$ and $m \times m$ matrices, where $m \leq n$. $\boldsymbol{B}$ is called a compression of $\boldsymbol{A}$ if there exists an orthogonal projection $\boldsymbol{P}$ onto a subspace of dimension $m$ such that $\boldsymbol{PAP} = \boldsymbol{B}$. The Cauchy interlacing theorem states that If the eigenvalues of $\boldsymbol{A}$ are $\lambda_1(\boldsymbol{A}) \leq \cdots \leq \lambda_n(\boldsymbol{A})$, and those of $\boldsymbol{B}$ are $\lambda_1(\boldsymbol{B}) \leq \cdots \leq \lambda_m(\boldsymbol{B})$, then for all $j$,

$$\lambda_j(\boldsymbol{A}) \leq \lambda_j(\boldsymbol{B}) \leq \lambda_{n-m+j}(\boldsymbol{A}).$$

Notice that, when $n - m = 1$, we have

$$\lambda_j(\boldsymbol{A}) \leq \lambda_j(\boldsymbol{B}) \leq \lambda_{j+1}(\boldsymbol{A}).$$

## APPENDIX B

### DEFINITION OF VECTORS $\boldsymbol{\alpha}_i^{(p)}$ & $\boldsymbol{\beta}_i^{(p)}$

The vectors $\boldsymbol{\alpha}_i^{(p)}$ and $\boldsymbol{\beta}_i^{(p)}$ for $p = 1, \dots, B$ and $i = 1, \dots, \boldsymbol{m}_p$

are $M_B \times 1$ and $(1 + M_B) \times 1$ column vectors, respectively. These vectors are defined as follows.

$$\boldsymbol{\alpha}_1^{(p)}(j) = \begin{cases} 1 & j = 1 + M_{p-1} \\ 0 & Otherwise \end{cases} \text{ for } p = 1, \dots, B,$$

$$\boldsymbol{\alpha}_i^{(p)}(j) = \begin{cases} -1 & j = i - 1 + M_{p-1} \\ 1 & j = i + M_{p-1} \\ 0 & Otherwise \end{cases} \text{ for } \begin{array}{l} p = 1, \dots, B \\ i = 2, \dots, m_p \end{array},$$

$$\boldsymbol{\beta}_1^{(p)}(j) = \begin{cases} -\sqrt{n_p} & j = 1 \\ 1 & j = 2 + M_{p-1} \\ 0 & Otherwise \end{cases} \text{ for } p = 1, \dots, B,$$

$$\boldsymbol{\beta}_i^{(p)}(j) = \begin{cases} -1 & j = i + M_{p-1} \\ 1 & j = i + 1 + M_{p-1} \\ 0 & Otherwise \end{cases} \text{ for } \begin{array}{l} p = 1, \dots, B \\ i = 2, \dots, m_p \end{array},$$

where $M_p = \sum_{i=1}^{p} m_i$ for $p = 1, \dots, B$ and $M_0 = 0$.

APPENDIX C

MAXIMUM DEGREE, METROPOLIS-HASTING & BEST CONSTANT WEIGHTING METHODS

The Metropolis-Hastings weighting method is defined as:

$$\boldsymbol{W}_{i,j} = \begin{cases} 1/(1 + max(d_i, d_j)) & j \in N_i, i \neq j \\ 1 - \sum_{j \in N_i} \boldsymbol{W}_{i,j} & i = j \\ 0 & otherwise \end{cases}$$

where $d_i$ and $d_j$ are the degrees of nodes $i$ and $j$, respectively and $N_i$ is the set of immediate neighbors of node $i$.

The Maximum degree weighting method is defined as:

$$\boldsymbol{W}_{i,j} = \begin{cases} 1/\max_k(d_k) & j \in N_i, i \neq j \\ 1 - d_i/\max_k(d_k) & i = j \\ 0 & otherwise. \end{cases}$$

The Best constant weighting method is defined as:

$$\boldsymbol{W}_{i,j} = \begin{cases} \alpha & j \in N_i, i \neq j \\ 1 - d_i \alpha & i = j \\ 0 & otherwise \end{cases}$$

In [9] it has been shown that the optimum choice of $\alpha$ for best constant weighting method is $\alpha^* = 2/(\lambda_1(\boldsymbol{L}) + \lambda_{n-1}(\boldsymbol{L}))$ where $\lambda_i(\boldsymbol{L})$ denotes the $i$-th largest eigenvalue of $\boldsymbol{L}$ and $\boldsymbol{L}$ is the Laplacian matrix defined as $\boldsymbol{L} = \boldsymbol{A}\boldsymbol{A}^T$ with $\boldsymbol{A}$ as the adjacency matrix of the sensor network's connectivity graph.